\documentclass[9pt,conference]{IEEEtran}
\usepackage{amssymb,amsthm,amsmath,array}
\usepackage{graphicx}
\usepackage[caption=false,font=footnotesize]{subfig}
\usepackage{xspace}
\usepackage[sort&compress, numbers]{natbib}
\usepackage{stmaryrd}
\usepackage{xcolor}
\usepackage{mathtools}
\usepackage{float}
\usepackage{textcomp}

\begin{document}
\title{Space Debris: Are Deep Learning-based Image Enhancements part of the Solution? }
\author{\IEEEauthorblockN{
        Michele Jamrozik, 
        Vincent Gaudilli\`ere,
        Mohamed Adel Musallam
        and 
        Djamila Aouada
        \\ 
    }
    \IEEEauthorblockA{
Interdisciplinary Center for Security Reliability and Trust (SnT) \\ University of Luxembourg  
        }
}
\maketitle
\begin{abstract}
The volume of space debris currently orbiting the Earth is reaching an unsustainable level at an accelerated pace.   The detection, tracking, identification, and differentiation between orbit-defined, registered spacecraft, and rogue/inactive space ``objects'', is critical to asset protection. The primary objective of this work is to investigate the validity of Deep Neural Network (DNN) solutions to overcome the limitations and image artefacts most prevalent when captured with monocular cameras in the visible light spectrum.  In this work, a hybrid UNet-ResNet34 Deep Learning (DL) architecture pre-trained on the ImageNet dataset, is developed. Image degradations addressed include blurring, exposure issues, poor contrast, and noise. The shortage of space-generated data suitable for supervised DL is also addressed. A visual comparison between the URes34P model developed in this work and the existing state of the art in deep learning image enhancement methods, relevant to images captured in space, is presented. Based upon visual inspection, it is determined that our UNet model is capable of correcting for space-related image degradations and merits further investigation to reduce its computational complexity. 





\end{abstract}

\section{Introduction}
There are currently more than 5,500 active satellites in orbit and existing plans to launch large-scale constellations into Low Earth Orbit (LEO), e.g., Starlink, OneWeb, Kuiper etc., suggest that number could exceed 58,000 by 2030 \cite{toomanysats}. These numbers are alarming! Even at existing orbital debris levels, the risk of catastrophic chain reaction collisions, described as the Kessler effect \cite{kessler1978collision}, is hugely significant. Consequently, efforts are currently underway to develop debris removal and collision avoidance systems \cite{mark2019review, zhao2020survey}. However, without the ability to accurately detect, track, and classify/identify space objects, attempts to ``safely'' remove or manage space debris, through whatever means, will ultimately fail. Therefore, interest in developing monocular space-borne pose estimation systems, designed to enable Space Situational Awareness (SSA) of uncooperative spacecraft/objects, and debris removal is growing \cite{kisantal2020satellite}. 

Unlike active sensors and stereo cameras, monocular optical image payloads do not require repeated and computationally expensive, in-situ calibrations, have less mass, lower complexity, and reduced power requirements \cite{cassinis2019review}.   However, optical images captured in space with monocular cameras come with their own set of complications. In particular, image quality is compromised due to the extreme conditions related to lighting, temperature, space weather, atmosphere, and motion found in the space image
capture environment. For example, lighting from the sun may cause over-exposure and the loss of detail in highlighted areas of an image. Conversely, shadowed portions of a space object may receive almost no light resulting in image underexposure. Hence, captures often exhibit poor contrast, i.e., important details required to assess the activity status and risks posed by an object, are not adequately represented in an image. Additional artefacts include noise (thermal or sensor), blurring due to relative orbital motion (Kepler) or exposure settings, and wavelength (color) shifts due to Doppler effects. 

Until now, the enhancement of images captured in space as a pre-processing step for improved classification has not yet been widely explored. Therefore, methods to enhance images with degradations similar those found in monocular space images were reviewed. Classical approaches to image enhancement are discussed in \cite{SurveyClassicalIE}.  DL approaches to low light image enhancement are found in \cite{LLIE, LvAttentionGuidedLowlight2020, 710815} and \cite{KinDPlus}. DL models addressing blurring and colour shifting  are found in \cite{Moran_2020_CVPR,yang2021laffnet, WangUnderwaterColorSpace}. UNet variations \cite{afifi2021learning, Yide_UNetplusplus, ChenDark, wang2021uformer} and Retinex \cite{Chen2018Retinex, cnn_survey} are popular DL architectures for image enhancement. A survey of DL approaches to image enhancement that includes the comparison models investigated in this work may be found in  \cite{LoLi_survey}.  

The primary goal of this work is to further improve upon existing space image enhancement models \cite{jamrozik2022image,ortiz2021deep} to automatically correct for degradations most commonly found in images captured in-space, i.e., poor contrast, blurring, and noise corruption to enable SSA and effective debris removal. 

\section{Methods}
In this investigation, monocular visible light images were investigated because they are more readily available and more cost effective to obtain than alternative payload images, e.g., infrared, hyperspectral, stereo, etc. In addition, comparison with existing DL image enhancement techniques can be more fairly conducted. Furthermore, and unlike thermal or stereo images, calibration is not required, thereby reducing the computational complexity of the method. 

 A variation of the original UNet architecture proposed in \cite{RonnebergerUNetConvolutionalNetworks2015} was chosen for this work for several reasons.  Firstly, UNet is a relatively simple architecture and reducing complexity is critically important for space imaging applications.  Next, the UNet architecture has a special symmetry and location information is built back into an image as it goes up the expansion side of the network through skip connections \cite{oyedotun2021training} present in the \cite{milesial} architecture version. The UNet DL architecture is, therefore, well suited for ``image-to-image'' applications like image enhancement. In addition, UNets may be trained with relatively small datasets which is highly relevant to the current application given the limited availability of ``real'' space image data.  Also, once trained, inference with UNet models is relatively fast. Finally, UNet models perform well on segmentation tasks which is a logical extension of the current work to facilitate pose estimation and SSA.  

Pixel-by-pixel loss function values such as mean-squared error, can be deceivingly large for noisy images. Therefore, the perceptual loss function described in \cite{johnson2016perceptual} which is better suited to the reconstruction of edges and fine details, was adopted in this work.  Our model, entitled URes34P, replaces the encoder side of a UNet with a 34 layer ResNet \cite{ResNet} pretrained on the  ImageNet dataset. In addition, it incorporates the INCR method to fill in new pixel values generated during the upsampling process required in multiple architecture layers \cite{pixelshuffle2016}.  Due to memory constraints, UNet architectures favor smaller batch sizes, especially for large image sizes \cite{RonnebergerUNetConvolutionalNetworks2015}.
 
Both compute capabilities and training time were limited for this research. To counter these restrictions, the network training methodology included the following measures: transfer learning, progressive resizing, one fit cycle, learning rate reduction on plateau, save best model, and early stopping. Additionally, batch normalization \cite{ioffe2015batch} and Adam optimization are employed \cite{Adam}.
SPARK, a synthetic dataset, \cite{SPARK2021} is the primary data source for this work.  

\subsection{Data Augmentation}
The URes34P model developed in this work is supervised. Supervised learning means that access to many input-target image pairs, exemplifying desired image properties, is required. However, access to ``real'', in-situ, space image-target datasets, is extremely limited. Therefore, space image data is  often simulated to overcome data shortages. To this end, the SPARK dataset was augmented with degradations including blurring, overexposure, underexposure, and noise population implemented via Albumentations \cite{albumentations}, and were added to copies of target images to be representative of those often found in space image data. Multiple image variations were generated from a single source image and share the same ground-truth/target image.

\section{Results \& Discussion}
\begin{figure}[ht!]
\centering
\includegraphics[width=.5\textwidth]{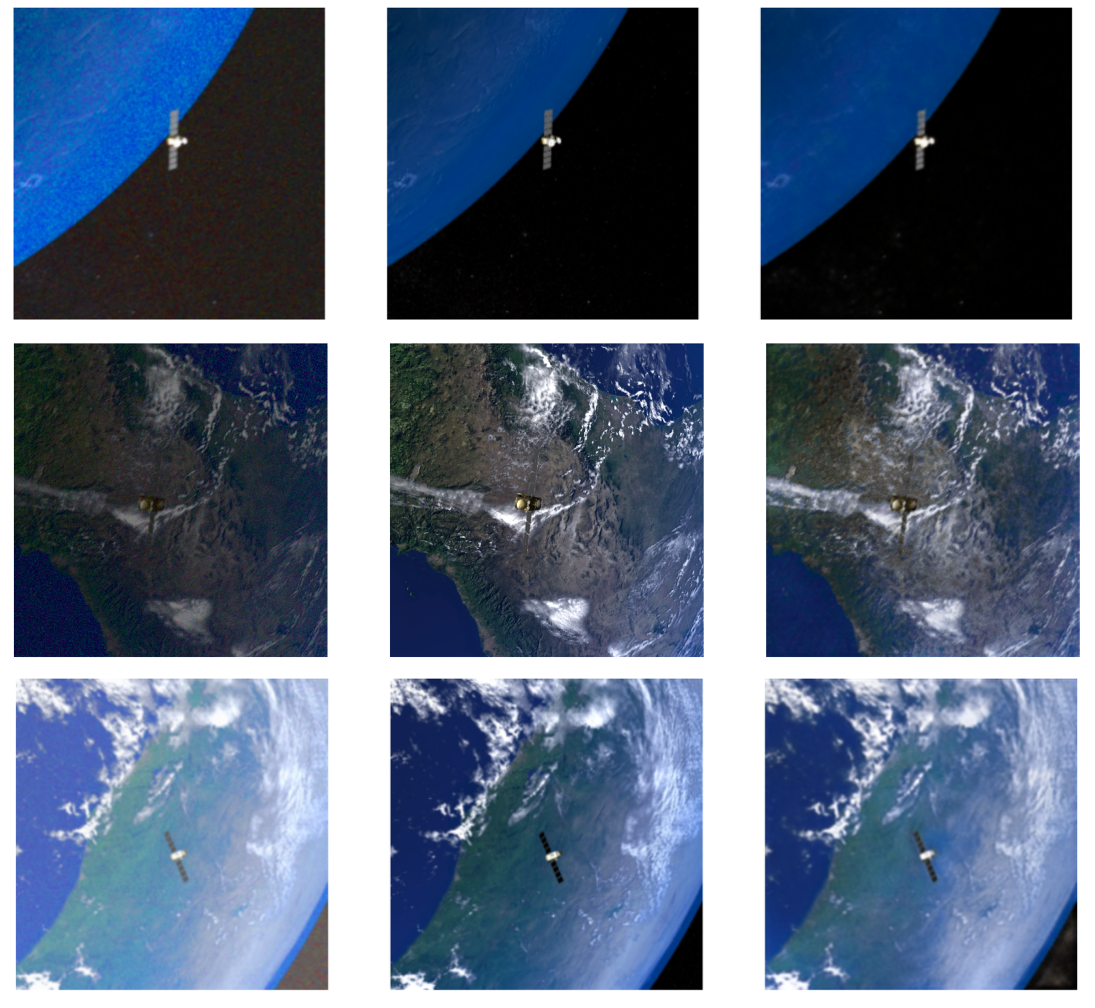}
\caption{URes34P model inference results obtained when tested with augmented images from the SPARK dataset. From left to right: input images, target images, enhanced results.}
\label{fig:DLcompare}
\end{figure}

\begin{figure}[!ht]
\centering
\includegraphics[width=.5\textwidth]{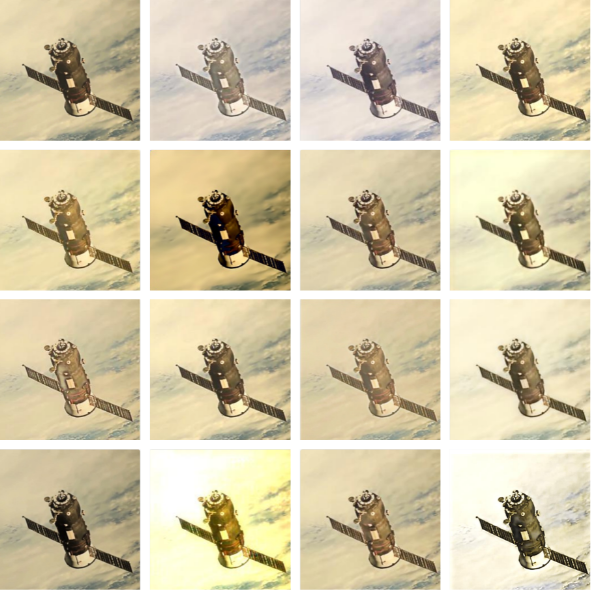}
\caption{A visual comparison of URes34P (bottom right corner) to existing DL image enhancement solutions referenced in \cite{LoLiPlatform}. From top left to bottom right: Input, Zero-DCE, Zero-DCE++, LightenNet, EnlightenGAN, MBLLEN, KinD, DRBN, Retinex-Net, RRDNet, TBEFN, LLNet, ExCNet, DSLR, KinD++, URes34P.}
\label{fig:DLcompare_space}
\end{figure}

A representative example of degraded SPARK dataset images enhanced via the URes34P model are shown in Fig.~\ref{fig:DLcompare}. Fig.~\ref{fig:DLcompare_space} shows the results of presenting ``real" images, captured in space, to URes34P and compared against existing DL image enhancement methods. According to visual results,  URes34P reduces image artefacts that impede accurate pose estimation including blurring and noise. Image contrast is also improved in tested images. The URes34P architecture developed in this work is similar to the Dynamic UNet model proposed in \cite{FASTAI_API} but was generated with different datasets, hyperparameters, and training protocol. Based upon visual inspection, the ability of URes34P to generalize beyond the training dataset justifies its further development to enhance non-simulated images to facilitate pose estimation and improve SSA. Additional loss functions and spacecraft power/energy saving optimization techniques such as pruning, quantization, and Neural Architecture Search (NAS) \cite{surianarayanan2023survey} merit further investigation.  In addition, extended training with more non-synthetic imagery should be considered to further improve generalization. 

\section{Conclusion}
The underlying premise of this work was to address whether the application of DL to pre-process images captured in space with modest payloads could be useful in circumventing and mitigating the effects of space object collisions. Based upon investigatory results, the answer to the question, ``Space Debris: Is Deep Learning-based Image Enhancement a Part of the Solution?'', is yes.\footnote{Acknowledgements:
This work was partly funded by the Luxembourg National Research Fund (FNR), under the project reference BRIDGES2020/IS/14755859/MEET-A/Aouada.} The ability of URes34P to generalize beyond the training dataset is confirmed and justifies its further development to enhance space images captured in-situ. Furthermore, the utility of incorporating DL into a space image enhancement pipeline, to facilitate pose estimation and SSA, is warranted. Future efforts will focus on improving results while reducing model complexity and speeding inference. 

\bibliographystyle{IEEEtran}
\bibliography{bibs/bib}
\end{document}